\documentclass{article}

\usepackage{PRIMEarxiv}

\usepackage[utf8]{inputenc} 
\usepackage[T1]{fontenc}    
\usepackage{hyperref}       
\usepackage{url}            
\usepackage{booktabs}       
\usepackage{amsfonts}       
\usepackage{nicefrac}       
\usepackage{microtype}      
\usepackage{lipsum}
\usepackage{fancyhdr}       
\usepackage{graphicx}       
\graphicspath{{media/}}     

\usepackage{amssymb}
\usepackage{amsmath}
\usepackage{amsthm}
\usepackage{mathtools}

\newtheorem{theorem}{\bf Theorem}[section]

\newtheorem{definition}{\bf Definition}[section]

\title{Graphs that predict exciton delocalization}

\author{
Gregory D. Scholes}

\author{
	Gregory D. Scholes \\
	Department of Chemistry \\
	Princeton University \\
	Princeton, New Jersey, 08540 U.S.A.\\
	\texttt{gscholes@princeton.edu} \\
}
\date{\today}

\begin{document}
\maketitle

\begin{abstract}
The field of molecular excitons and related supramolecular systems has largely focused on aggregates where nearest-neighbour couplings dominate. We propose that radically different states can be produced by moving beyond that paradigm. In practice, how to accomplish this task remains an open challenge because it requires development of ways to couple networks molecules more densely. In the present work we motivate why it would be worthwhile. We describe a merger of work developed in the field of discrete mathematics with concepts and needs for the field of molecular excitons. We discuss the reasons for exciton localization and posit how systems where the spectrum contains a gap can be robust to disorder, and thus maintain coherence, or delocalization. We propose that certain kinds of structures (expander graphs) specifying how molecules are coupled to each other, show such a gap and thus resilience to decoherence. We review relevant background known from graph theory. This perspective suggests a fascinating scope of new properties possible by demonstrating expander graph inspired excitonics. 
\end{abstract}


\newpage


\section{Introduction}

Delocalized molecular exciton states are observed in multichromophoric systems when the electronic coupling between the subsystems exceeds the spectral line broadening\cite{Scholes2006, ScholesQIS, excitondeloc1, excitondeloc2, Fidder1991, Bardeen2014, Scholes-Faraday}. These delocalized states are eigenstates of the system that indicate how we cannot distinguish excitations localized on any one chromophore of the many-chromophore system when the chromophores interact with each so as to exchange excitations. As a result, the appropriate eigenstates of the system comprise linear combinations of the possible basis states. The electronic coupling that exchanges excitations is mainly due to Coulombic coupling between transition densities\cite{Scholes2003} of the subsystems and is controlled by the position and orientation of chromophores relative to each other. 

The delocalized eigenstates of excitons are evidenced by shifts and splittings of the electronic absorption spectra. However, those spectral features are often hidden by line broadening. Line broadening comes from a combination of slow fluctuations of the transition energies of the molecular subsystems—we call that disorder—and dynamic fluctuations caused by coupling to thermal fluctuations, phonons, vibrations, etc—we call that homogeneous line broadening. Line broadening does more than simply hide the exciton states. The kinds of interactions between the chromophores and the environment (or bath) that produce line broadening also cause exciton states to be significantly more localized than one might expect\cite{Kuhn2015, Coherence2017, Tempelaar2013, Neaton2023, Scholes-Faraday}. The mechanism of exciton localization is reviewed below.  

Excitonic delocalization is not limited to span only two molecules, but may extend over several molecules, at least, if the electronic coupling is large enough compared to spectral line broadening. We have learned a lot about the interplay of line broadening and electronic coupling by studying photosynthetic light-harvesting complexes\cite{Mirkovic2017, FlemingQRB}. One well-studied case is the B850 band of LH2 from purple photosynthetic bacteria, which comprises 18 (depending on the species of bacteria) bacteriochlorophyll-\textit{a} chromophores arranged in a ring. These excitons are interesting because the electronic couplings between adjacent molecules are similar in magnitude to the standard deviation of the site energy disorder. Different kinds of experiments concluded different delocalization lengths. For example, analysis of circular dichroism data suggested that excitons are delocalized throughout the ring upon absorption of light, whereas superradiance indicated an equilibrium delocalization length of 2-4 molecules. Pullerits, et al.\cite{Pullerits1996} used pump-probe spectroscopy to estimate delocalization as a function of time by comparing the intensities of the excited state absorption and bleach bands. They found evidence for a more delocalized initial state that localized somewhat with time. Recent theoretical studies highlight the interplay between electronic coupling and dynamical site energy disorder, and how that interplay influences exciton delocalization and diffusion\cite{Excitondiff1, Excitondiff2, Excitondiff3}. 

How many molecules can excitation feasibly delocalize over? A complementary question arises: can we design chemical systems that are more resistant to this decoherence that localizes the state? If so, what are possible design strategies? One obvious (brute force) way forward is simply to design systems where the molecules interact more strongly. We could also minimize disorder by producing a pristine local environment for the chromophores, and reduce homogeneous line broadening by lowering the temperature. Disorder-free polydiacetyelene chains that are synthesized in a pure crystal of monomers thus show extraordinary coherent delocalization of the exciton along the conjugated backbone\cite{Yamagata2011}, estimated to be ~30–50 nm at 15K. These approaches work to some extent, but still allow only relatively short coherence lengths\cite{Scholes-Faraday}. Here we discuss alternative strategies for increasing excitation delocalization lengths in very complex multichromophore systems.. 

The present paper presents an overview of how the way components in a network are connected can have a profound effect on the states and their properties. The paper combines some of our ideas explored in recent papers\cite{Scholes-Faraday, Scholes2020, ScholesEntropy, ScholesQLstates} and some relevant background in algrebraic graph theory that has inspired the work. In particular, our recent papers have explored how structuring the interactions among entities (e.g. chromophores) and including many interactions between each entities and the others enables remarkable properties to be engineered into a dominant \emph{emergent} state.  An emergent state is seen in the spectrum as an eigenvalue (the largest or smallest in the spectrum) that clearly splits away from all the states in the spectrum (i.e. a gap develops). Emergent states tend to become more robust as the size of the system, that is, the number of interacting entities, increases.Owing to the spectral gap between the emergent state and the other states, it is unaffected by diagonal energy disorder (static or dynamic), unless that disorder is large compared to the gap. We have furthermore shown that the emergent state is incredibly robust to disorder in the structure of the graph\cite{ScholesEntropy}, so these systems do not need to be precisely engineered. The present paper introduce these ideas to researchers in the field of physical chemistry and excitonics, while explaining some of the underlying principles more carefully and introducing some new results, such as consideration of random lifts for graph construction.

The present paper focuses on graph theory for interacting systems. The concepts are entirely abstract, and might be more interesting in cases of networks that can be wired together with high connectivity, typically difficult to achieve for molecules, because the connections---the Coulombic coupling---are attenuated with distance. But, nevertheless, how might the graph theory framework  relate to excitonics? The main idea is to think about whether there are connectivity structures in very complex multi-molecular that control the spectrum (the eigenvalues of the bare system) in such a way that we expect lessening of the complicated interplay of electronic and disorder that localizes these states by, essentially, mixing the eigenstates through  energy fluctuations that dynamically interchange the order of eigenvalues.  Thus, we look for structures where the lowest energy eigenvalue is well-separated from all other eigenvalues, with the separation exceeding the other energy scales of static disorder and reorganization energy. Note that in the plots of spectra shown in the following sections this isolated eigenvalue will be shown as the \emph{largest} eigenvalue. That is because we plot the spectrum of the graph adjacency matrix (\emph{vide infra}). An isomorphic coupling matrix could simply take the off-diagonal elements to be negative, then the emergent eigenvalue is the \emph{lowest} eigenvalue.

\section{Localization versus Delocalization of Molecular Excitons}
The excitonic eigenstates of a perfect system form a ladder of states in the spectrum. That spectrum is decided by the structure of the Hamiltonian matrix, mainly by the off-diagonal terms that indicate which pairs of sites couple to each other. We usually put the disorder coming from line broadening into the distribution of diagonal (excitation) energies, and average over different manifestations of this disorder. However, for the purposes of the present paper we set all those diagonal energies to zero. We also set any non-zero off-diagonal coupling to 1. This setup, termed the \emph{adjacency matrix}, allows us to focus on the `wiring diagram' of couplings. In recent studies we have investigated averaging over variations of the `wiring diagram', thus introducing a kind of off-diagonal disorder coming from the network structure.

Coming back to the spectrum of exciton states, what causes them to become localized? To understand this, we need to describe the state that we measure using conditions of the experiment, such as spectral filtering. The state is written as a density matrix. The density matrix comprises an average, at the amplitude level, over the distribution of pure states that contribute to the result of the measurement. An important consequence of that average is that quantum correlations disappear unless the phases of each state in the distribution are more-or-less in phase. What the various kinds of disorder tends to do is increase the variety of exciton states (amplitudes and phases) that contribute to the average. The  phase information that locks several sites in step gets lost, leaving us with only short-range coherences---localized excitons.

The relevant density matrix for a system with a finite number, $N$, of basis states can be constructed as follows. We average over a sequence of $M$ measurements, where each measurement gives us information about an eigenstate $|\Psi_j \rangle$ for system $j$ in the ensemble:
\begin{equation}
	\rho = \sum_{j=1}^M p_j |\Psi_j \rangle \langle \Psi_j |
\end{equation}
where $p_j$ is the probability of measuring $|\Psi_j \rangle$. The density matrix is written in terms of the expansion of $|\Psi_j \rangle$ in the chosen basis
\begin{equation}
	|\Psi_j \rangle = c_{1j} \psi_{1} + c_{2j} \psi_{2} + \dots + c_{Nj} \psi_{N}
\end{equation}
where $c_{mj}$ are complex coefficients. Then, assuming our measurement solely resolves states $j$ (e.g. the lowest energy eigenstate or those that carry the oscillator strength), the $mn$ entry of the density matrix is
\begin{equation}
	\rho_{mn} = \sum_{j=1}^M p_j c_{mj}c_{nj}^*.
\end{equation}

Considering each eigenstate $|\Psi_j \rangle $ in our ensemble in turn, we construct its contribution to the density matrix, multiply by the probability $p_j$ and add that to the accumulating ensemble average. If we consider $M$ eigenstates selected at random from the ensemble, then every $p_j = 1/M$. Sometimes our measurement might probe a specific energy window or interval, so then we only include in our average those eigenstates whose eigenvalues lie in that window.

The off-diagonal values in the density matrix (written in an appropriate basis) indicate how well the eigenstate phases are correlated within the ensemble. The wavefunction coefficients for the eigenstate corresponding to our detection window show a lot of variation if disorder juggles the ordering and relative phase of the eigenstates.

We can show how excitons localize by averaging over the phases of various overlapping eigenstates with an example. First, we define a useful way to quantify delocalization. Although a more restricted measure than others defined on the density matrix, the Inverse Participation Ratio (IPR) is frequently and conveniently employed as a measure of delocalization\cite{Fidder1991, Scholes-Faraday, JangMennucci}. This measure, a second order statistical moment, looks at the variance of probabilities within a wavefunction   delocalized over $n$ sites, averaged over the ensemble (which is indicated by the angle brackets):

\begin{align}\label{1.1}
	\textrm{IPR} =  \Big[ \langle \sum_{j=1}^{n} |c_{mj} |^4 \rangle \Big]^{-1} ,
\end{align}

The IPR can be used in calculations based on wavefunctions. For example, when exciton wavefunctions are calculated as a function of static disorder by Monte-Carlo averaging. The crucial distinction here is that each wavefunction in the ensemble is calculated and the IPR is evaluated for each of these wavefunctions then statistically averaged. Aside from the special case of static disorder, that averaging is different than first averaging the wavefunctions themselves, by constructing a density matrix, then calculating the delocalization length. The IPR is a measure for the delocalization length\cite{Scholes-Faraday}.  As an aside, it is not so easy to include dynamic disorder, that is, homogeneous line broadening, in these kinds of calculations because the temporal interplay of electronic coupling and fluctuations is difficult to handle\cite{Yang2005, May2002, Chernyak1996}. 

\begin{figure}
	\centering\includegraphics[width=5.0in]{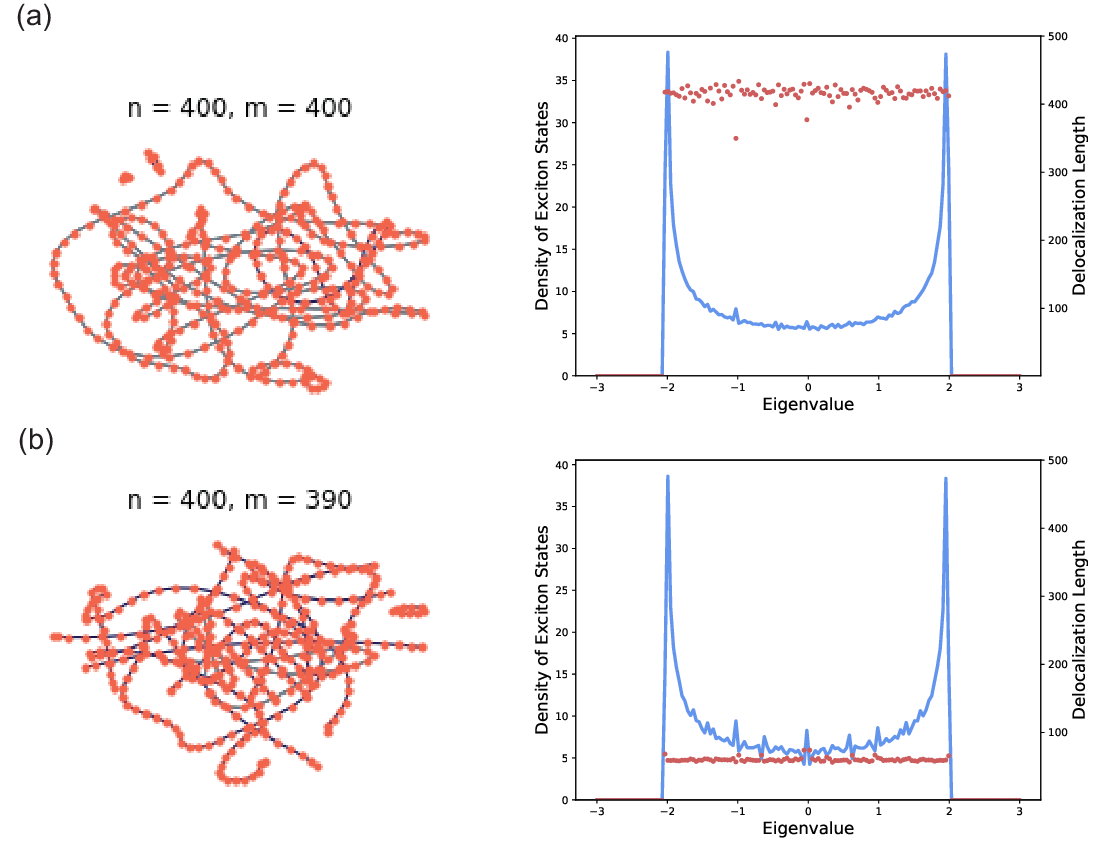}
	\caption{ (a) A random graph comprising only cycles, and therefore only nearest-neighbour coupling. The picture shows a representative graph on 400 vertices (with 400 edges). The results of an ensemble calculation of the density of exciton states for an ensemble of similar graphs, but with 900 vertices and 900 edges are shown to the right. The uncorrected delocalization length estimated from the IPR is shown by red dots. Since the calculation is for nearest-neighbour coupling, we should correct this by multiplying by 3/2. (b) As above, but 10 edges have been deleted randomly from the representative graph on 400 vertices that is drawn, giving 390 edges). 20 edges have been randomly deleted from each 900 vertex graphs that are used for the calculations (each now having 880 edges).  }
	\label{fig_1}
\end{figure}

Often we consider models for exciton states where molecules are coupled to their nearest neighbours. That is justified because the electronic coupling tends to fall off steeply with distance between the molecules, $R$, often as $1/R^3$. A typical example is where each molecule interacts with at most two neighbours. Let's consider a randomly constructed system of $n$ interacting molecules based on this blueprint, Fig ~\ref{fig_1}. For the calculations we have chosen 900 vertices. The pictures show smaller graphs (400 vertices) because they are easier to see. We first consider an ensemble of randomly generated structures where every molecule interacts with two others---meaning that the interaction patterns form cycles, top panel of Fig ~\ref{fig_1}a. The characteristic spectrum of the density of exciton states is shown together with the inverse of the IPR. The exciton is not delocalized over all 900 molecules partly because the ensemble comprises cycles of various sizes and also because the states are averaged in bins determined by the spectral resolution, or window, of our detection scheme. 

Now notice what happens when we randomly remove 20 edges from the total possible 900 edges. That is, we end up cutting open cycles to form some chains, Fig ~\ref{fig_1}b. Now the delocalization length is dramatically reduced because of this additional disorder. The main reason that this effect is so marked is that there is a high density of eigenstates crammed into a small spectral window. In fact, the spectrum of every one of these systems is contained in the interval $[-2, 2]$, no matter how many vertices the graph contains. This bound on the spectrum is specific to nearest-neighbour graphs. Thus, the larger we make our molecular aggregate (the larger is $n$), the higher the density of states in any small energy interval. This close energy separation of many states makes it the easier it is for line broadening to average over the phases of eigenstates in the spectrum and cause localization---smaller energy fluctuations cause greater disruptions in the ordering of eigenstates. Or, from another perspective, the resolution window of our measurement apparatus averages over more eigenstates of each aggregate. For this reason, very large delocalization length is challenging to achieve for systems where electronic coupling is predominantly nearest-neighbour, no matter how large is the electronic coupling. 

We conclude by proposing the rule that the more states we average over in our detection window, the more localized the exciton will be. In earlier work we noticed that, for certain patterns of coupling between molecules, we find spectra where the lowest (or highest, depending on the signs of the edges) eigenvalue is separated from the rest of the spectrum. We found that those states tend to be resilient to decoherence\cite{Scholes2020}. Therefore, to delocalize excitons we need to find patterns for coupling molecules together that allow states to `push away' from each other. In other words, a spectral gap can protect delocalization of states. To start with, we explain some background on how to describe and study coupling patterns systematically.

\section{Graphs as Coupling Maps}
It is convenient to represent the way molecules (or other entities) are coupled to each other using a graph, where the vertices denote the molecules and edges link molecules that are coupled, for instance by a sufficiently large Coulombic coupling. A graph $G(n,m)$, that we often write simply as $G$, comprises $n$ vertices and a set of $m$ edges that connect pairs of vertices. The size of a graph or subgraph, that is, the number of vertices, is written $|G|$. Rather than assigning a specific number of edges, sometimes it is convenient to indicate that  edges between vertices are found with probability $p$, so the graph is formally $G(n,p)$. Here we discuss only graphs with no loops and no multiple edges. For background see \cite{Diestel,Janson2000, Bollobas2001}. Occasionally in this paper we mention \emph{chromatic number} of a graph. If we colour each vertex of a graph so that no two adjacent vertices (i.e. vertices connected by an edge) have the same colour, then the chromatic number of the graph, $\chi(G)$ is the smallest number of colours required to colour all vertices of $G$ in this way. For the present paper the concept of chromatic number does not have special significance. 

The adjacency matrix of a graph $A$ is the $n \times n$ matrix containing entries $a_{ij} = a_{ji} = 1$, with $i, j \in \{ 1, 2, \dots n\}$, when an \textit{undirected} edge joins vertices $i$ and $j$. We may consider also \textit{signed} graphs, where the edge entry in the adjacency matrix is designated from $\{ 0, 1, -1 \}$ or \textit{weighted} graphs, where the edge entry in the adjacency matrix is scaled so that some edges carry more weight than others. In this paper we consider only normal undirected graphs, but it turns out that the results for other kinds of graphs are qualitatively comparable. That was evaluated by the author using extensive numerical studies (unpublished), while the following references will be helpful for the interested reader\cite{Mehatari, Reff2012, Zaslavsky1, Zaslavsky2}.

\begin{figure}
	\centering\includegraphics[width=5.0in]{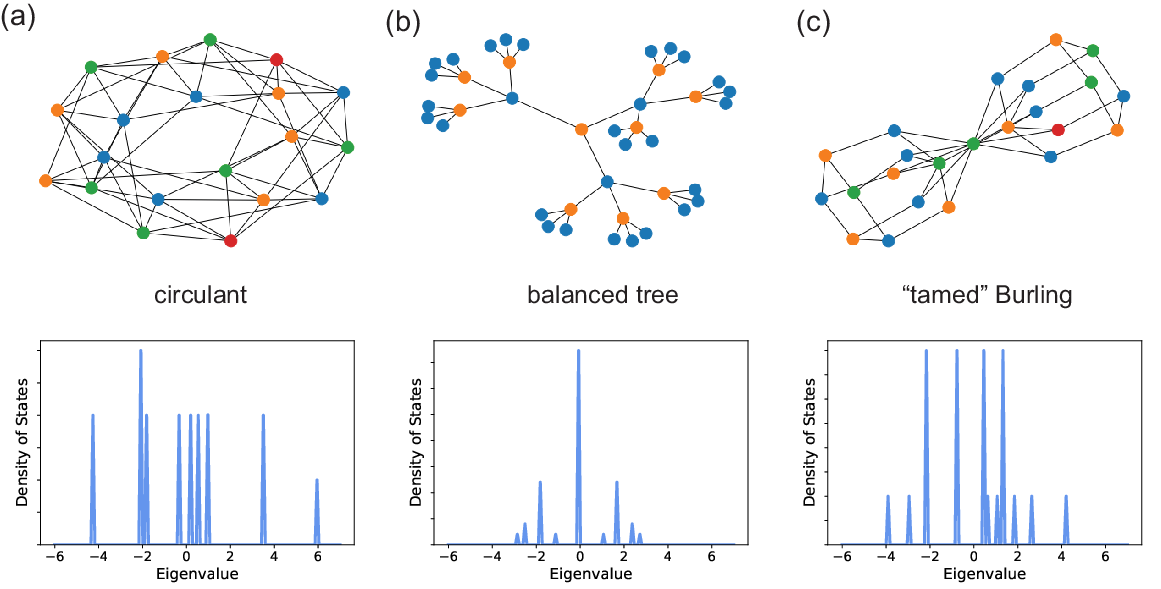}
	\caption{ Examples of graphs and their corresponding spectra. The vertices are coloured so that no pair of connected vertices have the same colour. The chromatic number of the graph is the minimum number of colours required to colour all vertices in this way. (a) Circulant graph. (b) Balanced tree. (c) Un unusual graph, the so-called `tamed Burling' graph. }
	\label{fig_2}
\end{figure}

The spectrum of a graph $G$ is defined as the spectrum (i.e. eigenvalues in the case of a finite graph) of its adjacency matrix $A$. Here we use the spectrum to detect and characterize states by their eigenvalues. The spectrum can tell us a lot about the graph, and \textit{vice versa}, see for example \cite{graphevals}. Examples of some small graphs are shown in Fig ~\ref{fig_2}.  The circulant graph on $n$ vertices has edges from each vertex $i$ to vertex $(i+x)\mod n$ and $(i-x)\mod n$, where $x \in {1, 2, 5}$. 

The `tamed' Burling graph is constructed according to the process described by Bonnet and co-workers\cite{Trotignon2024}. A good introduction to Burling graphs is given in ref \cite{Trotignon2023}. Burling graphs are triangle-free graphs constructed systematically in a sequence such that the chromatic number increases at each iteration. The graph shown in Fig ~\ref{fig_2}c is based on the third graph in the Burling sequence. To make the corresponding `tamed' Burling graph\cite{Trotignon2024}, or \emph{twincut} graph, the Burling graph is appended with `leaves' joined by an edge to each vertex at the end of the tree, here these are the vertices in the two 5-vertex cycles. Each leaf is then joined to the other vertices of the tree. This is, admittedly, a rather specialized graph for our purposes. Burling (in his thesis) developed the (untamed) graphs to investigate colouring of families of polytopes in Euclidian space. The tamed Burling graphs were developed to study fundamental questions about chromatic number and clique number in triangle-free graphs. 

A graph can also be used as a template to construct an ensemble of random matrices, whose spectrum is analyzed\cite{Scholes2020}. We mean `template' in the following sense. We typically want to produce an ensemble of matrices where the diagonal entries contain the frequency (or energy) of each site or entity (e.g. molecule, oscillator, etc). The frequency of each site in each matrix in the ensemble is randomly selected from a Gaussian distribution with appropriate mean and standard deviation. Generally, we do not add additional disorder to the couplings between each site. Instead, we assign them a value and enter them into off-diagonal positions in the matrix where the adjacency matrix is not zero.  Each matrix is diagonalized and we can accumulate the spectrum as a distribution, in addition to other quantities such as IPR or the density matrix.  This method is valuable for gauging how much the emergent state is affected by disorder with various standard deviations compared to the values of the coupling elements\cite{Scholes2020}. 

As mentioned in the Introduction, the main idea is to think about whether there are connectivity structures in very complex multi-molecular that control the spectrum (the eigenvalues of the bare system) in such a way that we expect lessening of the complicated interplay of electronic and disorder that localizes these states by, essentially, mixing the eigenstates through  energy fluctuations that dynamically interchange the order of eigenvalues.  Thus, we look for structures where the lowest energy eigenvalue is well-separated from all other eigenvalues, with the separation exceeding the other energy scales of static disorder and reorganization energy. In general, the ideas of the present paper are useful when the emergent state splits away from the other states by an energy that exceeds the variance of the static and dynamic disorder. To quantify this statement, below we focus much discussion on $d$-regular graphs. The gap between the emergent state and the second eigenvalue is $(d - 2\sqrt{d-1}) \times V$, where $V$ is the electronic coupling (assumed uniform) between vertices. Let's say $V = 50$ cm$^{-1}$ and $d = 10$, then the gap is $200$ cm$^{-1}$. It gets more interesting (and more impractical for molecules) with larger values for $d$. For instance, put $d = 101$, then the gap becomes $4050$ cm$^{-1}$ and the emergent state could involve hundreds of molecules.

\section{Expander Graphs and Emergent States}
\begin{figure}
	\centering\includegraphics[width=5.0in]{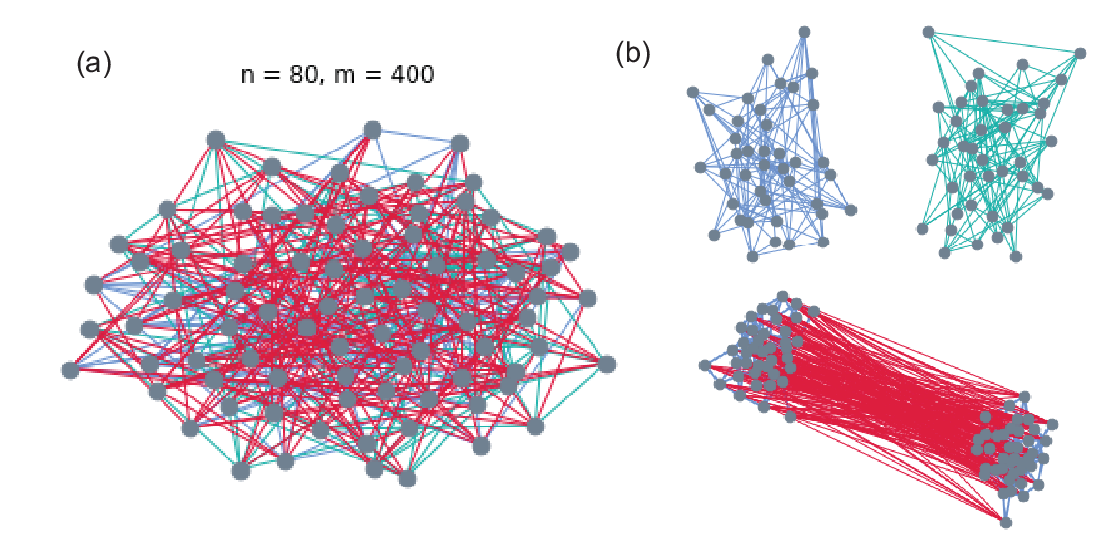}
	\caption{ (a) A  $d$-regular random graph on 80 vertices with $d = 10$. (b) The vertex set of this graph has been arbitrarily partitioned into two sets. The top panel shows these two sets and the edges within each set. The lower panel shows the edges connecting vertices from from set to the other, that is, the boundary of each set. }
	\label{fig_3}
\end{figure}

We proposed above that to protect a state in an open system from decoherence we should isolate it from the other eigenstates of the system. It turns out that there are families of graphs that do just this---expander graphs\cite{Sarnak2004, expandersguide, Lubotzky, Expanders, Expanders2, Alon1986}. Expander graphs are highly connected graphs that are optimal for communications networks. The advantage they provide is a guaranteed short pathway connecting any pair of vertices. While an expander graph could have only three edges from each vertex, they are usually are much more highly connected than a typical molecular system. An example is drawn in Fig ~\ref{fig_3}. Connectivity throughout the graph is quantified by the isoperimetric constant.

\begin{definition}
	(Isoperimetric constant) The isoperimetric constant of a graph $G$ with vertex set $V$, is defined as
	\begin{equation}
		h(G) = \min \Big\{ \frac{|\partial Y|}{|Y|}  \Big\} 
	\end{equation}
	with $Y \subset V$ and $|Y| \le \tfrac{1}{2}|V|$ . Here, $\partial Y$ (think of this as a single symbol, nothing to do with partial derivatives) is the boundary of $Y$, which means the set of edges in $G$ that have one endpoint in $Y$ and one endpoint in $V \setminus Y$. 
\end{definition}

In Fig ~\ref{fig_3} we show an arbitrary selection of vertices $Y$ and $V \setminus Y$ from a graph. The edges within these sets are coloured blue and green respectively. The edges connecting these sets ($\partial Y$) are coloured red. The graph drawn in this example is an expander graph, so no matter how we select the two sets of vertices, there will be many red edges connecting the sets. The concept of isoperimetric constant allows us to define an expander graph by considering families of graphs that are highly connected, no matter how large we make them (i.e. how many vertices they contain). These are known as expander graphs.

\begin{definition}
	(Expander families) Let $d$ be a positive integer. Let $(G_n)$ be a sequence of $d$-regular graphs on $n$ vertices such that $|G_n| \rightarrow \infty$ as $n \rightarrow \infty$. We say that $(G_n)$  is an expander family if the sequence $(h(G_n))$ is bounded away from zero.
\end{definition}

A well studied example of an expander family are the $d$-regular graphs, where every vertex is adjacent to $d$ other vertices in the graph.

\begin{definition}
	($d$-regular graph) A graph $G$ is $d$-regular if every vertex has degree (valency) $d$. That is, every vertex connects to $d$ edges.
\end{definition}

The special case of $d = 2$, that is cycles, like those considered in Fig ~\ref{fig_1}, are not expander graphs. To see that, consider what happens to $h(G)$ as a cycle becomes infinitely large. The physical picture of connectivity measured by the isoperimetric constant can be illustrated by arbitrarily dividing the vertices of  $d$-regular graph into two sets, then pulling those sets apart, Fig ~\ref{fig_3}. Notice the large number of edges that stretch out between these vertex sets.

The concept of a Cayley graph allows systematic construction of graphs, like the $d$-regular expanders, using group operations and thereby can simplify explanations of the properties of these graphs. It is also possible to define \emph{random} Cayley graphs, as described in, for example, \cite{Alon1994}, although we do not need those details here.

\begin{definition}
	(Cayley graph) Let $\Gamma$ be a group and $S \subset \Gamma$. The Cayley graph with respect to $S$, $\text{Cay}(\Gamma, S)$ is defined as follows. The vertices of $\text{Cay}(\Gamma, S)$ are the elements of $\Gamma$. Two vertices, $u, v \in \Gamma$ are adjacent if and only if there exists $s \in S$ such that $u = vs$ (i.e. $v^{-1}u \in \Gamma$).
\end{definition}

As an example, a 4-regular graph on $n$ vertices can be specified by $\Gamma = \mathbb{Z}_n$ and $S$ being a set elements of $\Gamma$ (excluding 0 and including the inverse of each element). The group operation is addition.  Recall that $ \mathbb{Z}_n$ means integers modulo $n$. For example, let's assume that our graph has $n = 8$, that is, 8 vertices, and let's put $S = \{1, 3, 5, 7 \}$. The edges are determined by taking each vertex at a time, then applying each of the group operations to it. Label the vertices $0, 1, 2, \dots 7$. Thus, vertex 1 connects to $1 + s$, where $s \in S$, meaning the vertices $0, 2, 4, 6$. 

Therefore, the graph $\text{Cay}(\Gamma, S)$ we produce is controlled by the generating set $S$ and the graph is $|S|$-regular, where $|S|$ means the size of the set $S$. Now the group characters decide the eigenvectors of the adjacency matrix of $\text{Cay}(\Gamma, S)$. The eigenvalue of the so-called trivial character, the operation that sends every element to 1, equals $|S|$ (equivalently the degree of each vertex in the graph). That is, $\lambda_0 = d$ for a $d$-regular graph. The eigenvectors are $f_j = (1, \exp(2\pi ij/|G|)), \exp(2\times 2\pi ij/|G|), \exp(3\times 2\pi ij/|G|), \dots/\sqrt{|G|}$, with $j = 0, 1, \dots, |G|-1$. The associated eigenvector is So, for $j = 0$, we have  $f_0 = (1, 1, 1, \dots)/\sqrt{n}$, if the graph as $n$ vertices. This is the emergent state, and is the reason that the largest eigenvalue of a $d$-regular graph equals $d$. The reader may also like to work this result out, and add some detail too, using the Perron-Frobenius theorem for eigenvalues and eigenvectors of non-negative matrices. See Sec. 8.8 of \cite{GodsilRoyle} or Sec. 8.4 of \cite{Horn}. Incidentally, the Perron-Frobenius theorem is closely connected to principles used in the Coulson-Rushbrooke theorem\cite{CoulsonRushbrooke}.  

It turns out that there is a deep connection between the isoperimetric constant $h(G)$ of a graph $G$ and the graph's second-largest eigenvalue that we denote $\lambda_1(G)$, or simply $\lambda_1$. It is the first `non-trivial' eigenvalue. Recall that the eigenvalues of a graph are the eigenvalues of the graph's adjacency matrix. We order the eigenvlues $\lambda_0 \ge \lambda_1 \ge \dots \ge \lambda_{n-1}$. As discussed above, $\lambda_0$ is associated with the constant function on the vertices, that is the eigenvector $(1, 1, 1, \dots)\sqrt{n}$. It can be shown that if $G$ is a $d$-regular graph then the isoperimetric constant for the graph is bounded as follows:

\begin{equation}
	h(G) \le \sqrt{2d(d - \lambda_1)} .
\end{equation}

We sketch the proof of this relation here. The interested reader might refer to ref  \cite{expandersguide}, section 9 of chapter 1, for more details. The result is obtained by first bounding the spectral gap $\lambda_0 - \lambda_1 = d - \lambda_1$ (recall that $\lambda_0 = d$, as shown above). We use the Rayleigh-Ritz theorem\cite{expandersguide}, which is basically the minimax principle for eigenvalues in ref \cite{Bhatia},
\begin{equation}
	\langle \Delta f, f \rangle_2 \le d - \lambda_1(G),
\end{equation}
where $\Delta$ is the graph Laplacian operator that is constructed from the diagonal matrix with entries equal to the degree of each vertex, $D$, with the adjacency matrix $A$, $\Delta = D - A$. The subscript 2 indicates that we take the $L^2$-norm on the inner product. Each $f_i$ is a normalized eigenfunction of $A$ associated with eigenvalue $\lambda_i$. Note that, as evident from the definition, \{Laplacian eigenvalues\} = $d$ - \{adjacency eigenvalues\}. We can then establish that 
\begin{equation}
	\frac{h(G)^2}{2d} \le \langle \Delta f, f \rangle_2,
\end{equation}
by considering the properties of the Laplacian operator on the graph and establishing how edges across the graph are indicated in the eigenvectors. For details, see \cite{expandersguide}, section 9 of chapter 1.

The result, Eq (4.2) shows that expander graphs have a gap between the first two eigenvalues. For $d$-regular graphs, the gap can be substantial, and is optimal at the the Alon-Boppana bound\cite{Alon1986, Nilli1991}: 

\begin{theorem} (Alon-Boppana)
	Let $G$ be a $d$-regular graph, then
	\begin{equation}
		\lambda_1 \ge 2\sqrt{d - 1} - o_n(1).
	\end{equation}
\end{theorem}

A convenient way to construct $d$-regular graphs is to use Cayley graphs, where the vertices of the graph are all the elements of a group, $\Gamma$, and the edges are defined using the group operation with a subset of the group elements, $S$.

The setting of Cayley graphs (defined above) allows us to connect the isoperimetric constant and spectral gap to a third property of expander graphs known as weak containment\cite{KazhdanTbook, Lubotzky, UnitaryRepbook, Fell1962, Zuk2003}. By considering unitary representations $\pi(s)$ of each group operation $s$, we can examine how these operations act on vectors in a Hilbert space $\xi \in \mathcal{H}_{\pi}$ associated to $\Gamma$. The unitary representation $\pi$ of the group is said to contain `almost invariant' vectors if
\begin{equation}
	\lVert \pi(s) \xi - \xi \rVert < \epsilon,
\end{equation}
for all $s \in S$, where $\lVert \dots \rVert$ means norm (defined by the inner product). This quantity tells us how the vectors get `rotated away' from the emergent state vector by the group operations. For expander families, that rotation tends to a limit away from zero as graphs get larger, for every $\pi(s)$. We then say that the graphs have Kazhdan's property $(T)$. 

Property $(T)$ means that the trivial representation of $\Gamma$, that produces the eigenvalue $d$ for a $d$-regular graph (i.e. the emergent state), is `bounded away' from the other irreducible representations\cite{Lubotzky2005}. That is, we have a spectral gap. We define a Kazhdan constant $\kappa(\Gamma, S)$ by minimizing eq 4.6 over all irreducible, nontrivial, unitary representations of $\Gamma$. It can then be established that
\begin{equation}
	\kappa \ge \sqrt{\frac{2(d - \lambda_1)}{d}},
\end{equation}
where $\Gamma$ is a nontrivial group, $S \subset \Gamma$ and $d = |S|$. A spectral gap can therefore be used as sufficient condition to have property $(T)$ \cite{Zuk2003}. 

A very simple example will illustrate the concept of property $(T)$, and then show why 2-regular graphs (cycles) do not form an expander family. Let $\Gamma = \mathbb{Z}_n$ and $S = \{1\}$, so that $\text{Cay}(\Gamma, S)$ is the $n$-cycle. A matrix (a $1 \times 1$ matrix in this instance) representation of the group is just
\begin{equation}
	\rho_a(k) = \exp \Big(\frac{2\pi i ak}{n} \Big),
\end{equation}
which gives $n$ representations of $\mathbb{Z}_n$ as $\rho_0, \rho_1, \dots \rho_{n-1}$. We are interested in the non-trivial representations compared to the trivial representation (the identity representation, $\rho_0$). Define $\alpha = e^{i \theta}$, with $\theta = 2\pi/n$, then $|\alpha - 1| = 2 \sin \frac{\theta}{2}$, see Lemma 8.7 of ref \cite{expandersguide}. We interpret $|\alpha - 1|$ as the distance from $\alpha$ to 1 in the complex plane. This illustrates how the vectors are rotated away from the emergent state. For example, when $\Gamma = \mathbb{Z}_5$, $\rho_1(1) = \theta = 2\pi/n$ is a counterclockwise rotation by $72^{\circ}$. Now let $\xi$ be any unit vector in $\mathbb{C}$ (i.e., it lies on the unit circle), then we can find the minimum rotation, which gives the Kazhdan constant,
\begin{equation}
	\kappa(\Gamma, S) = \lVert \rho_1(1) \xi - \xi \rVert  = \lVert \rho_{n-1}(1) \xi - \xi \rVert =  |\alpha - 1| = 2\sin \frac{2\pi}{n}.
\end{equation}
Now, notice that for any $n$, $\kappa(\Gamma, S) > 0$, so the graph has property $(T)$. But, $\kappa(\Gamma, S) \rightarrow 0$ as $n \rightarrow \infty$, so the cycles do not form an expander family. 

\begin{figure}
	\centering\includegraphics[width=2.5in]{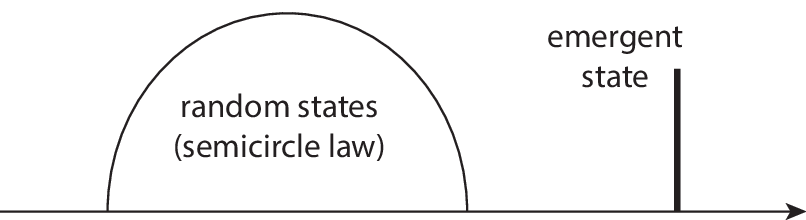}
	\caption{ Schematic drawing of the spectrum of a $d$-regular random graph. }
	\label{fig_4}
\end{figure}

Spectra of $d$-regular graphs on $n$ vertices comprise two distinct features: the emergent state with eigenvalue $d$ and a set of `random states' with eigenvalues in the interval $[-2\sqrt{d-1}, 2\sqrt{d-1}]$, Fig ~\ref{fig_4}. The random states are random in the sense that, for an ensemble  of large $d$-regular graphs, they are distributed according to Wigner's semicircle law for random matrices\cite{McKay1981}. This highlights the many connections between random graphs and random matrices. 

\begin{definition}
	(Random graph)  A random graph on $n$ vertices, $G(n,m)$, is generated by randomly assigning $m$ edges among pair of vertices. We can think of this graph as one of the $2^{n \choose 2}$ graphs in the set of all random graphs on $n$ vertices. The adjacency matrix of a random graph $G(n,m)$ is a symmetric random Bernoulli matrix---that is, a symmetric matrix containing $m$ random entries of 1. 
\end{definition}

\begin{definition}
	($d$-regular random graph)  The $d$-regular random graphs have similar properties to the subset of $d$-regular graphs. The $d$-regular random graphs are a subset of random graphs with the sole restriction that every vertex has degree $d$. 
\end{definition}

\begin{figure}
	\centering\includegraphics[width=5.0in]{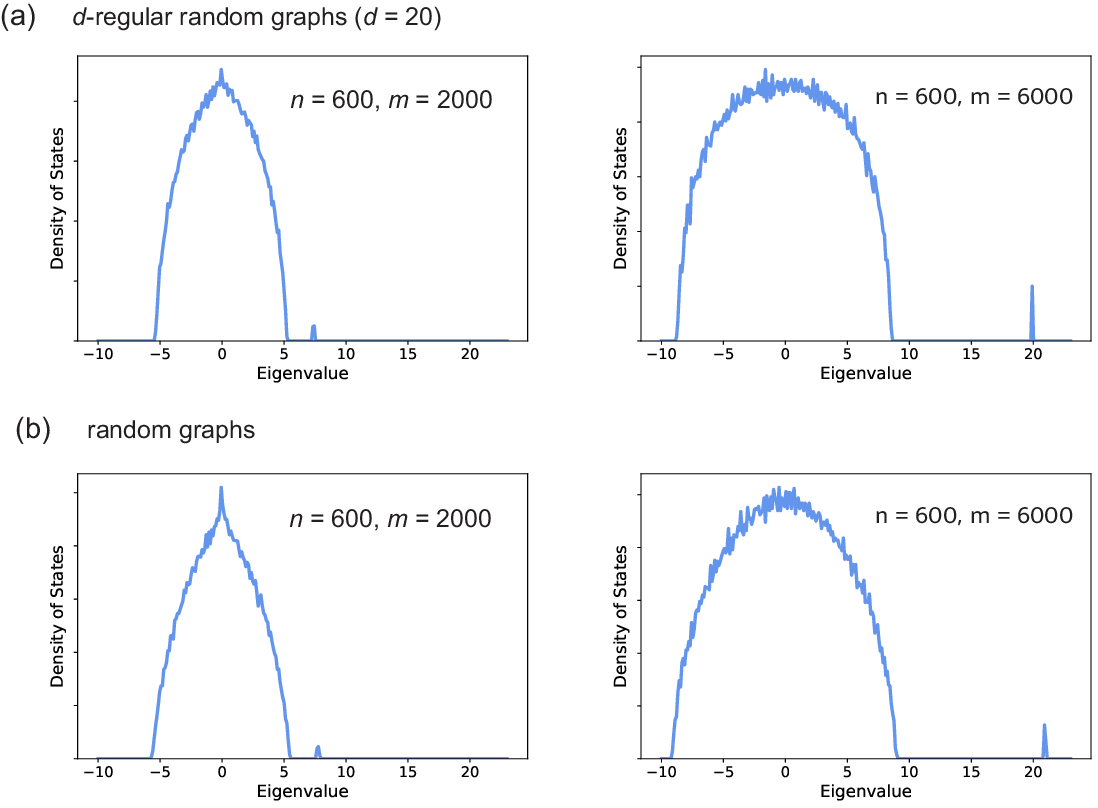}
	\caption{ (a) Examples of ensemble spectra of $d$-regular random graphs. (b) Examples of ensemble spectra of unrestricted random graphs. }
	\label{fig_5}
\end{figure}

In Fig ~\ref{fig_5} we compare ensemble spectra for random graphs on 600 vertices to $d$-regular random graphs, also on 600 vertices. The random graph is constructed by randomly adding edges (2000 or 6000 depending on the simulation) between vertices. There are no restrictions on how the edges are added. In the case of the $d$-regular graph, the construction is a little more involved. We produce a random $k$-regular graph, which means it has $600 k/2$ edges in total. Now we randomly delete edges until we are left with the required total  (2000 or 6000 depending on the simulation). Therefore, the graph, which is the product of a percolation, is no longer $k$-regular, but is $d$-regular, where $d < k$, and therefore the emergent eigenvalue does not lie at $k$. See ref \cite{ScholesEntropy}. We derive an estimate for the size of $d$ below.

It is interesting to note that the spectra for these finite graphs are not too different. That is because, in each case, the number of edges in the random graph exceeds the threshold for the `phase transition'. The phase transition in random graphs is one of the most important and pioneering results in the field\cite{Janson2000, Bollobas2001}. The idea is that there is a threshold on the number of edges in a random graph, so that below the threshold the graph comprises small islands of connected vertices, whereas above that threshold the graph is dominated by a single giant connected component. Therefore, the additional edges added after formation of the giant component increase the density of the graph---that is, they increase the average degree. 

The close connections between random graphs, random matrices, and $d$-regular (random) graphs gives an intuition for why the emergent state---despite its remarkable order (in the eigenstate)---is produced spontaneously by the collective effect of many (possibly small) interactions among sites at the vertices. The pattern of interactions is almost random. They are subject to one rule only, that the degree of each vertex is $d$. For a graph with many vertices, the entropic cost to produce the spectrum containing the emergent eigenvalue, overall, is counter-intuitively very small. Numerical results\cite{ScholesEntropy} show that the emergent state is robust to percolation. We can randomly remove  about half the edges and still retain the emergent state. In each example, the eigenvalue of the emergent state was found at $d$, where $d$ is the average degree of all vertices in the graph. We can justify that observation as follows. 

Guionnet describes how to analyze the spectrum of Bernoulli random matrices\cite{Guionnet}. We can adapt her derivation in the dense case to explain the spectrum of the disordered $k$-regular graphs (i.e. $k$-regular graphs with edges removed randomly such that the average degree is $d$). We start by summarizing the method for random graphs, where each adjacency matrix in an ensemble is a symmetric $n \times n$ Bernoulli matrix $\mathbf{B}_n$ with entries of zero on the diagonal. We can write this matrix as the matrix with all off-diagonal entries equal to $p$, so that the average degree of a vertex is $pn$. Now rewrite $\mathbf{B}_n$ in the form
\begin{equation}
	\mathbf{B}_n = \sqrt{np(1 - p)} \mathbf{X}_n + p\mathbb{I},
\end{equation}
where $\mathbb{I}$ is a matrix with all off-diagonal entries set to 1. Entries of $\mathbf{X}_n$ are centered and renormalized so that
\begin{equation}
	\mathbf{X}_n = \frac{\mathbf{B}_n - p}{\sqrt{np(1 - p)}}  .
\end{equation}

The matrix $p\mathbb{I}$ has one non-trivial eigenvalue of $\lambda^B_1 = pn$, equal to the average degree when the graph is large. By Weyl's interlacing properties\cite{Bhatia, GodsilRoyle}
\begin{equation*}
	\lambda^X_n \le \lambda^B_n \le \lambda^X_{n-1} \le \dots \le \lambda^X_1 \le \lambda^B_1 ,
\end{equation*}
we infer $\lambda^B_1$ is the largest eigenvalue. 

To adapt this method for $k$-regular graphs with random edge deletions, we should consider an ensemble of Bernoulli matrices that have each been masked (randomly) so that the off-diagonal entries containing $p$ appear only in the pattern of any $k$-regular graph. For each Bernoulli matrix in the ensemble we determine the corresponding $\mathbf{X}_n$, and thus rewrite $\mathbf{B}_n$, but now written in as  a sum of a term proportional to $\mathbf{X}_n$ and $p\mathbb{I}_d$, where $\mathbb{I}_d$ is a matrix containing entries of 1 in the shape of the mask. The latter term is the salient feature because, for every matrix in the ensemble, its nontrivial eigenvalue is found at $pk = d$, where $d$ is the average degree. That is, the emergent state of an ensemble of disordered $k$-regular graphs has eigenvalue equal to the average degree. 

\section{Graphs from Random Lifts}
Prior work has shown how random coverings, or lifts, of graphs can be used to produce expander graphs\cite{Linial2002, Linial2006, Spielman2015}. We can start with any graph and perform a sequence of random lifts to produce graphs that are more and more connected. Here we focus simply on 2-lifts\cite{Linial2006, Spielman2015}. For a graph $G$, the 2-lift $\hat{G}$ contains two vertices $v_0, v_1$ for each vertex $v$ in $G$. This pair is known as the \emph{fibre} of the vertex $v$. The 2-lift is the \emph{double cover} of $G$ if edges are assigned in $\hat{G}$ as follows. If $(u,v)$ is an edge in $G$, then $\hat{G}$ contains a pair of edges: $(u_0,v_1)$ and $(u_1,v_0)$.

A closely related operation can be accomplished by taking sequences of Mycielskians of a graph. The Mycielskian of $G$ is constructed from the double cover of $G$ by adding an extra vertex, as detailed below. Here we study sequences of Mycielskians of graphs.  

\begin{figure}
	\centering\includegraphics[width=5.0in]{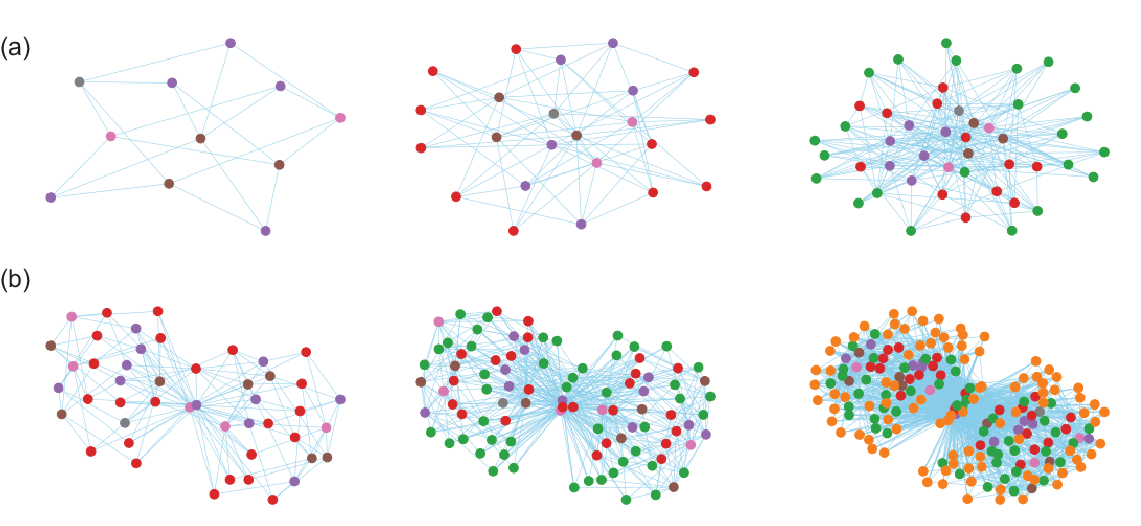}
	\caption{ (a) Mycielskian sequence starting from the cycle on five vertices ($C_5$). (b) Mycielskian sequence starting from the `tamed Burling' graph shown in Fig ~\ref{fig_2}. The vertices are explicitly coloured to illustrate the sequence of chromatic numbers. }
	\label{fig_6}
\end{figure}

Given a graph $G$ with vertex set $V(G) = {v_1, \dots, v_n}$, the Mycielskian graph\cite{Mycielski} $G_m$ of $G$ contains $G$ as a subgraph, together with $n+1$ additional vertices, ${u_1, \dots, u_n, w}$, giving a total of $2n+1$ vertices. Each vertex $u_i$ is the mirror of vertex $v_i$ in $G$ and $w$ is an extra vertex. Each vertex $u_i$ is connected by an edge to $w$. The other edges are decided as follows. For each connected vertex pair $v_i, v_j$ in $G$, the graph $G_m$ includes two edges, $u_i, v_j$ and $v_i, u_j$. The construction of these graphs was motivated by that fact that the chromatic number of $G_m$ is always equal to one more than the chromatic number of $G$. By iterating the Mycielskian construction, we can generate a sequence of graphs, $G, G_{m1}, G_{m2}, \dots$, where $G_{m(i+1)}$ is the Mycielskian graph of $G_{mi}$. Some examples are shown in Fig ~\ref{fig_6}.

\begin{figure}
	\centering\includegraphics[width=2.5in]{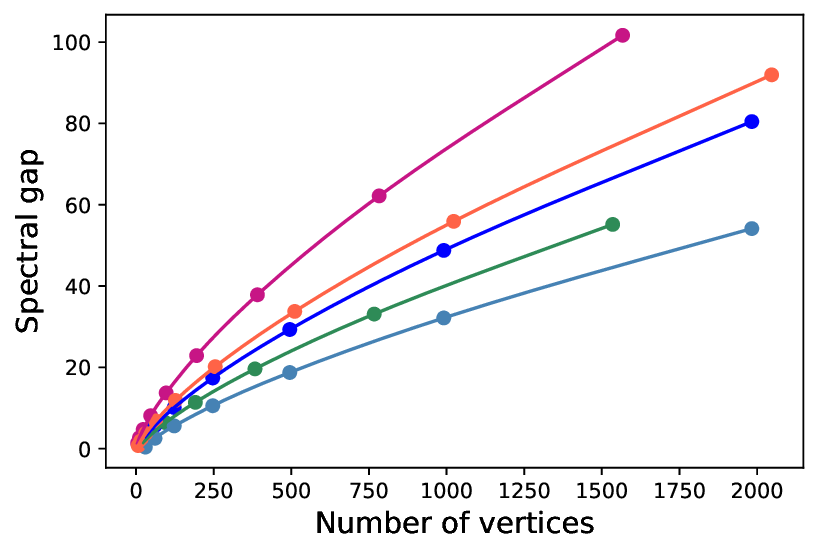}
	\caption{ Plots of the spectral gap (i.e. $\lambda_0 - \lambda_1$) of various Mycielskian sequences starting from the base graph $G$. From top to bottom the sequences start from: $C_4$, $C_7$, a 10-regular random graph, `tamed Burling' graph shown in Fig ~\ref{fig_2}, a 3-regular random graph. }
	\label{fig_7}
\end{figure}

In Fig ~\ref{fig_7} we show plots of the spectral gap (i.e. $\lambda_0 - \lambda_1$) of various Mycielskian sequences. It is seen that for sequences starting from $G$ being a cycle, that the spectral gap expansion is greatest for smaller cycles. We plot here, for example, the 4-cycle compared to the 7-cycle sequences.  In the case of $G$ being a small $d$-regular graph, then the sequence of spectral gaps of is larger when $d$ is larger. 

The Mycielskian sequences are not the same as sequences that define expander families, Def. 4.2. Nevertheless, the process clearly provides a systematic way to increase the spectral gap with each iteration. We also know that expander graphs with nearly optimal gap can be constructed using a series of 2-lift operations\cite{Linial2006}. Marcus and co-workers\cite{Spielman2015} further proved that this strategy can produce families of graphs that indeed have an optimal gap---the Ramanujan graphs\cite{Sarnak1988}. Also, note that each graph in the Mycielskian sequence has chromatic number equal to one more than the preceding graph in the sequence, that is, $\chi(G_m) = \chi(G_{m-1}) + 1$.



\section{Discussion}
We have described the properties of large systems comprising coupled nodes. These nodes would represent molecules, atoms, or even macroscopic systems. In the case of coupled molecules, the eigenstates are molecular  exciton states. The principle that we have put forward is that delocalization, or coherence, of the state of interest can be protected from decoherence by ensuring that the state is well separated in the spectrum from all the other states. This property is enabled by certain schemes for coupling the nodes, known as expander graphs Expander graphs are typified by the $d$-regular (random) graphs, but we showed examples of other constructions that also have the expander property.  Thus, there are many different blueprints for coupling nodes so that the emergent states is robust to various kinds of disorder. 

In the examples described, we use the convention that coupling terms are positive because we display spectra of the  graphs. However, the coupling can be negative, in which case the emergent state becomes the lowest energy state. Also, we did not consider energetic disorder in the node frequencies (or couplings) and we assume that fluctuations in the electronic couplings are small compared to the mean value. We did, however, include structural disorder by random graph constructions or by random edge deletion. Energy disorder in the node frequencies (diagonal disorder) is commonly considered in the analogous random matrix theory. We have addressed it in prior work\cite{Scholes2020}. Here we focused on the principles controlling emergence of the state. 

In recent work, we explored how expander graphs can be linked in ways that enable the emergent state to represent superposition states\cite{ScholesQLstates}. We called that the 'quantum-like' (QL) bit. These QL bits can be connected to other QL bits to produce arbitrary new states. The concept shows how the emergent state of an aggregate system can be made much richer.  That work opens up many questions. We explicitly discussed two major open questions: (i) Is there a `quantum-like' advantage for function? Perhaps this could be leveraged in hardware for QL-computing, which would be greatly simplified from that needed for true quantum computing, but may offer an attractive compromise between classical and quantum architectures. (ii) Could QL states and function be used by neural networks in the brain? We concluded that quantum-like states can exist in arbitrarily complex systems and seem to provide a concrete staring point for proposing testable hypotheses for quantum (-like) functions in very complex systems. Importantly, QL-states allow a way to overcome the objection for various proposals of quantum states in biology, the brain, and other complex systems---that is, that quantum states decohere quickly and thus cannot to serve a functional role. The key is to consider `quantum like' states\cite{Khrennikov1, Ozawa2020, KB2013, Khrennikov2016}, not quantum states. 

An obvious feature of expander graphs is that many edges (connections among the nodes) are required, and that the edges must span all length scales throughout the graph. This is what makes them so interesting. However, from the chemist's viewpoint, this property renders them difficult to fabricate. It is clear that we can construct macroscopic examples of expander networks by `wiring' oscillators together. So they can be manifest in electronic circuits or neural networks, for example. But how might we exhibit such structures 
on the molecular scale? 

The prototypical example of a molecular scale expander is the molecular polariton. Here tens of thousands, or more, molecules are coupled to an effective mode of the radiation field to produce a giant collective state. The representative graph is a star graph, where the center mode is the radiation field. Since this is a bipartite graph, we see two emergent states---the upper and lower polaritons. The molecular polariton phenomenon suggests how we might go about producing more sophisticated coupling structures among large numbers of molecules. We speculate that various kinds of dielectric patterning might be able to produce suitable coupling networks\cite{KG1988, Huber2017, Bennett2020}. Another strategy might be to shape correlations in the environment\cite{Bittner2024}. These proposals would likely benefit from devising effective graph structures using sparsification techniques\cite{Spielman2011}. 

To sum up, the field of molecular excitons and related supramolecular systems has largely focused on aggregates where nearest-neighbour couplings dominate. We propose that radically different states con be produced by moving beyond that paradigm; which remains an open challenge in practice. Here we presented a merger of work developed in the field of discrete mathematics with concepts and needs for the field of molecular excitons. This perspective suggests a fascinating scope of new properties possible by demonstrating expander graph inspired excitonics.


\section*{Acknowledgments}
	Noga Alon is thanked for suggesting the papers on random lifts. This material is based upon work supported by the National Science Foundation under Grant No. 2211326 and the Gordon and Betty Moore Foundation through Grant GBMF7114.

\bibliographystyle{unsrt}  
\bibliography{Scholes_5}


\end{document}